# BRINGING AI PIPELINES ONTO CLOUD-HPC: SETTING A BASELINE FOR ACCURACY OF COVID-19 AI DIAGNOSIS

Iacopo Colonnelli[1*], Barbara Cantalupo[1], Concetto Spampinato[2], Matteo Pennisi[2], Marco Aldinucci[1]

[1]*University of Torino, Computer Science Dept., Corso Svizzera 185, 10149, Torino, Italy*

[2]*University of Catania, Electrical Engineering Dept., Viale Andrea Doria 6, 95125, Catania, Italy*

**ABSTRACT.** HPC is an enabling platform for AI. The introduction of AI workloads in the HPC applications basket has non-trivial consequences both on the way of designing AI applications and on the way of providing HPC computing. This is the leitmotif of the convergence between HPC and AI. The formalized definition of AI pipelines is one of the milestones of HPC-AI convergence. If well conducted, it allows, on the one hand, to obtain portable and scalable applications. On the other hand, it is crucial for the reproducibility of scientific pipelines. In this work, we advocate the StreamFlow Workflow Management System as a crucial ingredient to define a parametric pipeline, called "CLAIRE COVID-19 Universal Pipeline", which is able to explore the optimization space of methods to classify COVID-19 lung lesions from CT scans, compare them for accuracy, and therefore set a performance baseline. The universal pipeline automatizes the training of many different Deep Neural Networks (DNNs) and many different hyperparameters. It, therefore, requires a massive computing power, which is found in traditional HPC infrastructure thanks to the portability-by-design of pipelines designed with StreamFlow. Using the universal pipeline, we identified a DNN reaching over 90% accuracy in detecting COVID-19 lesions in CT scans.

[*] Corresponding author. E-mail: *iacopo.colonnelli@unito.it*

# 1 Introduction

The ability of AI-related techniques to transform raw data into valuable knowledge is growing at a breakneck pace. Among these techniques, Deep Learning (DL) has benefited from crucial results in Machine Learning (ML) theory and the large availability of data. The accuracy of the process is strictly related to the quality and size of available datasets. Usually, more data means more accurate predictions, but also more computing power needed to train the model.

In particular, in the last decade, Deep Neural Networks (DNNs) became larger and larger, and nowadays, a reasonably sized DL workload cannot prescind from the availability of heterogeneous (GPU-equipped) computing resources. For this reason, High-Performance Computing (HPC) is undoubtedly an enabling platform for AI, and, in turn, AI enables success in scientific challenges where traditional HPC techniques have failed. For their part, supercomputers are shifting more and more to heterogeneous hardware, both because of their better energy efficiency and to satisfy the ever-increasing need for GPU-enabled workloads pushed by DL.

Despite this potential, supercomputers are rarely used for standard AI workloads. This is mainly due to technical barriers, i.e., user-unfriendly SSH-based remote shells and queue-based job submission mechanisms, which prevent AI researchers without a strong computer science background from effectively unlocking their computing power. In practice, HPC centres are not designed for general purpose applications. Only scalable and computationally demanding programs can effectively benefit from the massive amount of processing elements and the low-latency network interconnections that characterize HPC facilities, justifying the high development cost of HPC-enabled applications. Moreover, some seemingly trivial features are not supported by HPC facilities, e.g., exposing a public web interface for data visualization in an air-gapped worker node.

On the other hand, the technical barriers to Cloud-based infrastructures lowered substantially with the advent of the *-as-a-Service model. Alas, the multiple layers of virtualization that characterize modern cloud architectures introduce significant processing overheads and make it impossible to apply adaptive fine-tuning techniques based upon the underlying hardware technologies, making them incompatible with performance-critical HPC applications.

In this work, we advocate the combination of two distinct approaches as an effective way to lower the technical barriers of HPC facilities for AI researchers [1]:

- A *cluster-as-accelerator* design pattern, in which cluster nodes act as processing elements of user-defined tasks sent by a Cloud-based, general-purpose host executor. This paradigm can be used to offload computation to HPC facilities in a more intuitive way, as it mimics the GPGPU paradigm used by DL applications in a GPU-equipped machine;
- *Hybrid workflows*, i.e. workflows whose steps can be scheduled on independent and potentially not intercommunicating execution environments, as a programming paradigm to express this design pattern while ensuring portability and reproducibility of complex AI workloads.

In the evaluation part, we apply these two principles to a real AI application, i.e., a DNN training pipeline for COVID19 diagnosis from Computed Tomography (CT) scan images. The StreamFlow [2] Workflow Management System (WMS), which supports hybrid Cloud-HPC workflows as first-class citizens, is used as the underlying runtime system.

## 2 The StreamFlow toolkit

A workflow is commonly represented as an acyclic digraph $G = (N, E)$, where nodes refer to different portions of a complex program and edges encode *dependency relations* between nodes. In this representation, a direct edge connecting a node $m$ to a node $n$ means that $n$ must wait for $m$ to complete before starting its computation.

The workflow abstraction has already been explored for offloading computation to HPC facilities transparently [3]. Many of the existing WMSs come with a diverse set of *connectors*, some of them addressing Cloud environments and some others more HPC-oriented. Nevertheless, a far smaller percentage can deal with *hybrid workflows*, offering the possibility to seamlessly assign each portion of a complex application to the computing infrastructure that best suits its requirements.

Hybrid workflows can strongly reduce the necessary tradeoffs in relying on such high-level abstraction, both in terms of performance and costs. Indeed, complex applications usually alternate computation-intensive and highly parallelizable steps with sequential or non-compute-bound operations. When scheduling such applications to an HPC centre, only a subset of steps will effectively take advantage of all the available computing power, resulting in a low cost-benefit ratio.

Moreover, HPC facilities do not support some common operations that are instead trivial on Cloud-based infrastructures, such as exposing web interfaces for data visualization.

The StreamFlow toolkit[2] [2], whose logical stack is depicted in Fig.1, has been specifically developed to orchestrate hybrid workflows on top of heterogeneous and geographically distributed architectures. Written in Python 3, it supports the CWL coordination standard [4] for expressing workflow models through a declarative JSON or YAML syntax. The translation of these declarative semantics into an executable workflow model is delegated to `cwltool`, the CWL reference implementation. The StreamFlow runtime layer is then able to efficiently execute such a model by translating it into a dataflow graph, identifying independent steps and running them in parallel whenever possible.

---

[2]https://streamflow.di.unito.it/

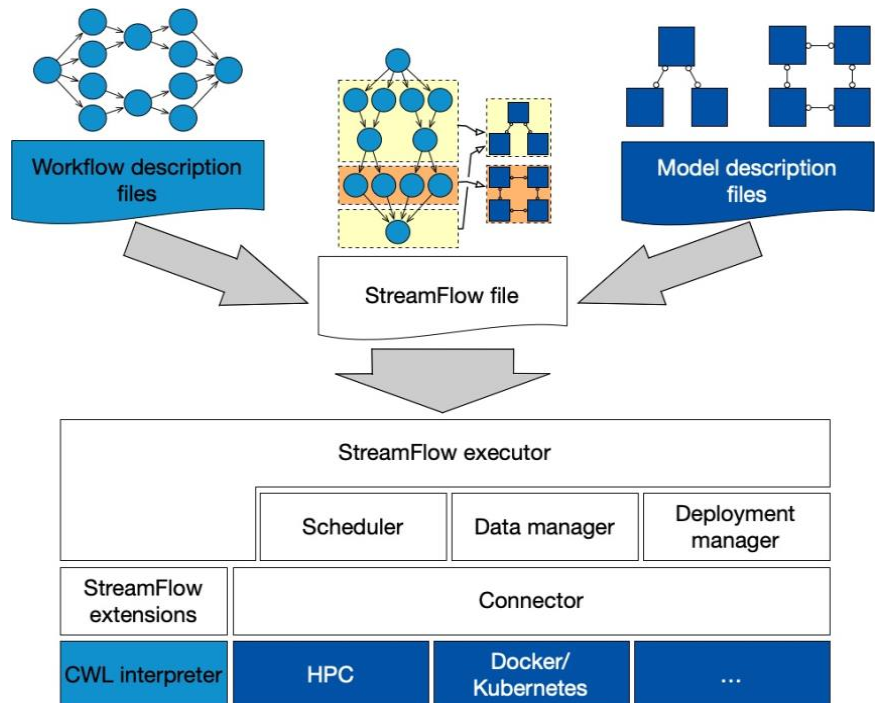

**Fig**.**1**: The StreamFlow toolkit's logical stack.

Alongside, one or more execution environments can be described in well-known external formats, e.g., Helm charts for Kubernetes deployments or Slurm files for HPC workloads. A `streamflow.yml` file constitutes the entry point of a StreamFlow run, relating each workflow step with the best suitable execution environment. This feature actually plugs the hybrid layer in the workflow design process.

Another distinctive feature of the StreamFlow WMS is the possibility to manage complex, multi-agent execution environments, ensuring the *co-allocation* of multiple heterogeneous processing elements to execute a single workflow step. The same interface can then be used to describe a diverse ecosystem of distributed applications, ranging from MPI clusters running on HPC facilities to microservices architectures deployed on Kubernetes. To provide enough flexibility, StreamFlow adopts a three-layered hierarchical representation of execution environments:

- A `model` is an entire multi-agent infrastructure and constitutes the *unit of deployment*, i.e., all its components are always co-allocated when executing a step;
- A `service` is a single agent in a model and constitutes the *unit of binding*, i.e., each step of a workflow can be offloaded to a single service for execution;
- A `resource` is a single instance of a potentially replicated service and constitutes the *unit of scheduling*, i.e., each step of a workflow is offloaded to a configurable number of service resources to be processed.

All communications and data transfer operations are started and managed by the StreamFlow controller, removing the need for bidirectional channels between the management infrastructure and the target resources and allowing tasks to be offloaded to HPC infrastructures with air-gapped worker nodes. Moreover, StreamFlow does not need any specific package or library to be installed on the target resources, other than the software dependencies required by the host application. As a consequence, virtually any target infrastructure reachable by a practitioner can serve as a target model, as long as a compatible connector implementation is available.

# 3 The CLAIRE COVID-19 universal pipeline

To demonstrate how StreamFlow can help bridge HPC and AI workloads, enabling reproducibility and portability across different platforms, we present the COVID-19 universal pipeline, developed by the Confederation of Laboratories for Artificial Intelligence Research in Europe (CLAIRE)[3] task force on AI & COVID-19 during the first COVID-19 outbreak. The group, composed of fifteen researchers in complementary disciplines (Radiomics, AI, and HPC) and led by Prof. Marco Aldinucci [5], investigated the diagnosis of COVID-19 pneumonia assisted by Artificial Intelligence (AI).

At the start of the pandemic, several studies outlined the effectiveness of chest radiology imaging for COVID-19 diagnosis. Even if X-Ray scans represent a cheaper and most effective solution for large scale screening, their low resolution led AI models to show lower accuracy than those obtained with CT scans. Therefore, the latter has become the gold standard for the investigation of lung diseases.

Several research groups worldwide began to develop DL models for the diagnosis of COVID-19, mainly in the form of deep Convolutional Neural Networks (CNN). As is especially the case in the medical field, reproducibility of the results was an important issue to address. Providing AI pipelines for COVID diagnosis with reproducible steps should not be an option to ensure the goodness of the results. This is particularly important when dealing with DL models, which are obscure by definition.

Furthermore, such a significant number of proposals was not accompanied by any baseline of the accuracy expectation. A comprehensive study of the proposed solutions highlighted that it would not be easy to evaluate the most promising approaches due to the adoption of different architectures, pipelines and datasets. Therefore, instead of proposing yet another hopefully better solution, the task force commitment was to organize the knowledge so far to consolidate and formalize all or the most state-of-the-art deep learning models to diagnose COVID-19.

The result of such commitment was the distillation of a reproducible workflow, the *CLAIRE COVID-19 universal pipeline* represented in Fig.2, capable of automating the comparison of the proposed DL models and supporting the definition of a baseline for any further evaluation.

The pipeline is basically designed by composing the main steps in a standard AI workflow. TC scan images are pre-processed to insulate the region of interest, the lungs in this case, and then used to train a classifier to recognize the typical lesions of interstitial pneumonia caused by COVID-19, specifically consolidation, crazy paving and grown glass.

The pipeline is composed of two main parts:
- A *data preparation* phase (yellow elements), comprising *pre-processing*, where standard techniques for cleaning the training images are applied, and *segmentation*, for extracting and selecting the region of interest for the next training. This step is performed just once for each dataset;
- The *core training* phase (blue elements), composed of standard AI steps such as *data augmentation*, to generate image variants, model *pre-training*, to generate an initial set of weights for initialization, and eventually *classification*, which labels each image with a class identified with a kind of lesion that is typical of the disease. The final steps are *cross-validation*, which increases the pipeline robustness by applying the training on different portions of the dataset, and *performance metrics*, obtained by collecting and comparing all the measures from the different pipelines.

The effectiveness of the DL approach depends on many parameters, e.g., the input dataset, the pre-processing steps, the chosen DL model, and the hyperparameters of the training algorithm, such as learning rate, weight decay, learning rate decay. It is worth noting that, in the universal pipeline, the DNN itself, which is the model used for classification, is just one of the variables that can be set for training. Different variants of existing networks (such as AlexNet, ResNet, DenseNet) can be plugged into the pipeline, but any future network could be included in principle.

---
[3] https://claire-ai.org/

Finally, the pipeline is modelled as a workflow where every single step is a kind of container without any dependency on external libraries or vendor-specific technology. This choice enables portability on different platforms allowing to run the same pipeline in different platforms or even across different ones.

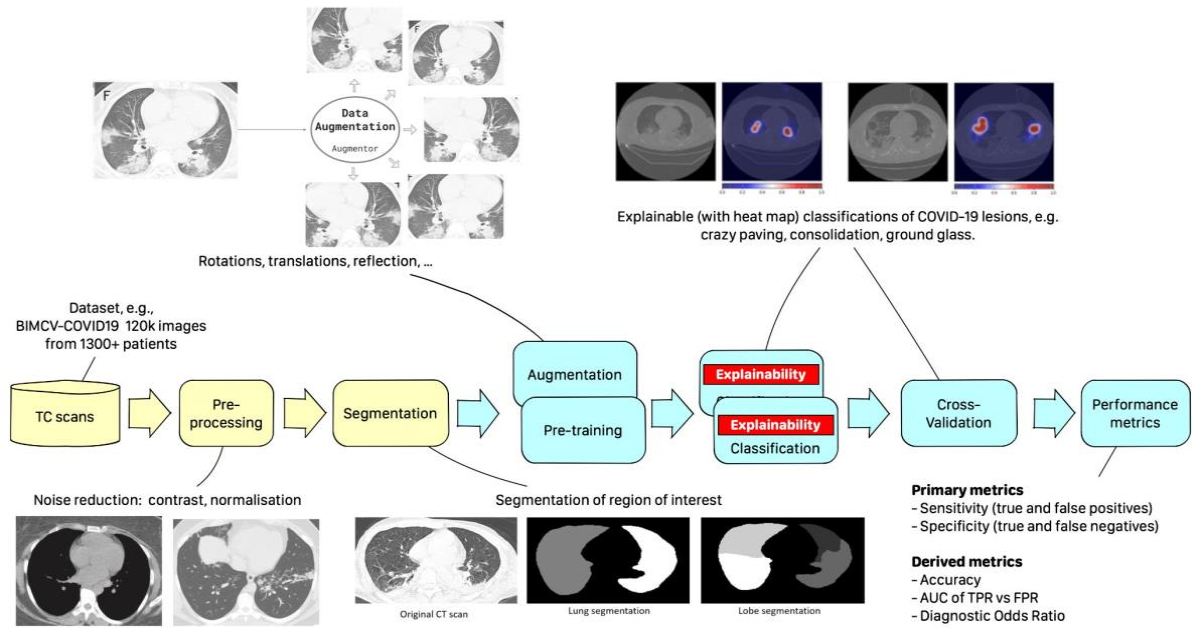

**Fig**.**2**: The CLAIRE COVID-19 universal pipeline.

The first set of experiments has been already completed comparing about 1% of the variants (11 of 990) and applying different segmentation types. Results show that the pipeline can generate models with excellent accuracy in classifying typical interstitial pneumonia lesions due to COVID-19, with sensitivity and specificity metrics over 90% in the best cases. More detailed information on the experiment's results are available in [6] and demonstrate that the proposed approach is able to carry out the same task with an accuracy that is at least on par with, or even higher than, human experts, thus showing the potential impact that these techniques may have in supporting physicians in decision making.

## 4 Experimental evaluation

On the move from the design to the implementation, the universal pipeline takes advantage of the StreamFlow technology. The workflow is defined using CWL, using traditional parallel computing operators (such as scatter, broadcast, gather, reduce) to explicitly annotate the parallelizable portions of the pipeline (Fig. 3). The search space is composed of all the combinations of network models, training hyperparameters, and one or more datasets. Each point in such space is a tuple, which constitutes the input for a single pipeline instance. As each pipeline instance is independent of each other, the overall execution is a typical embarrassingly parallel problem, whose parallelism can be exploited by distributing the input tuples to the work units through the "scatter" operator.

In turn, every single pipeline instance can further exploit parallelism, distributing different partitions of the dataset (generated by the cross-validation algorithm) to as many instances of the classification step through another scatter operator. In this setting, the initial weights of the DNN are broadcasted to all the classifiers of a specific pipeline instance. Performance metrics are then collected through a combination of reduce operators, first reducing internally in the single pipeline instance and then globally collecting results from all the pipelines. The baseline performance of the analyzed pipelines for COVID-19 diagnosis is finally obtained.

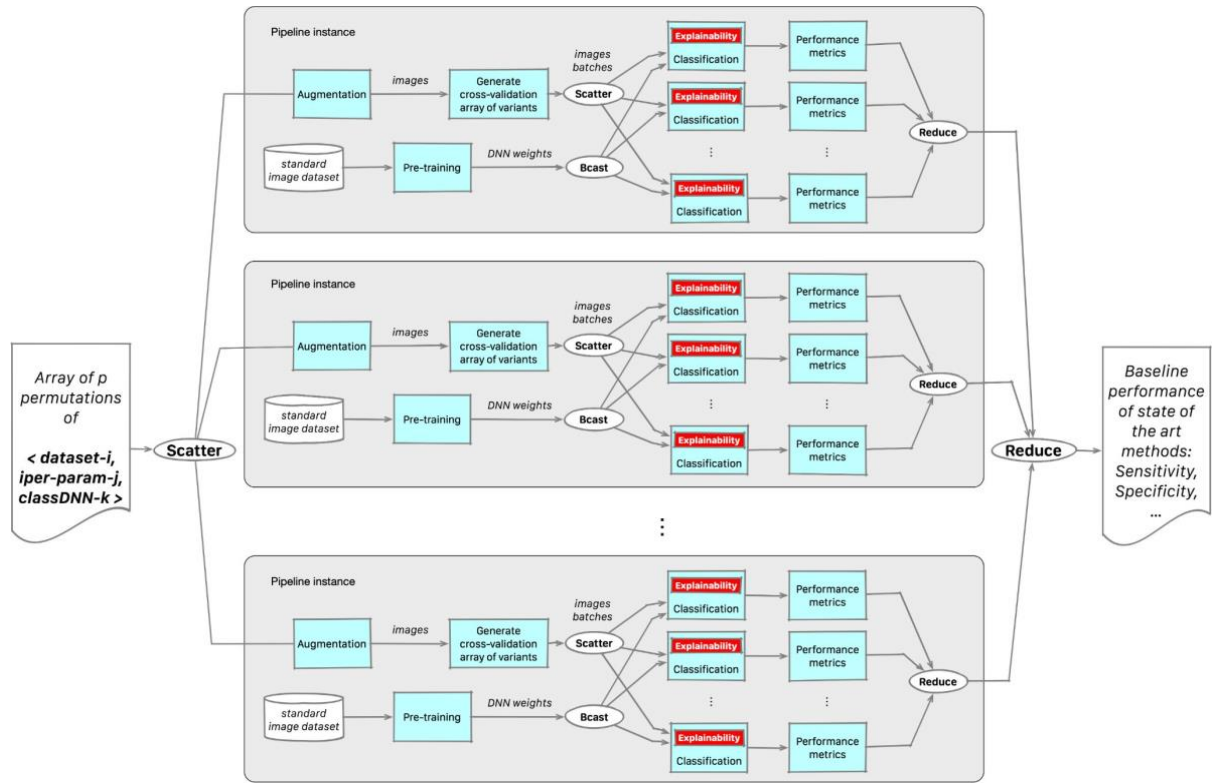

**Fig.3**: Unfolded implementation of the pipeline training components

For evaluation purposes, we ran the pipeline on the BIMCV-COVID19 dataset, with more than 120k images from 1300 patients. Assuming to train each pre-trained model for 20 epochs on such dataset, a single variant of the pipeline takes over 15 hours on a single NVidia V100 GPU, one of the most powerful accelerators in the market. Therefore, exploring all the 990 pipeline variants we have selected would take over two years. Fortunately, as we already pointed out, the universal pipeline has an embarrassingly parallel structure, and therefore using a supercomputer could reduce the execution time down to 15 hours in the best case (i.e., when 990 GPUs are available at the same time).

Post-training steps, as performance metrics extraction and comparison, are better suited for a Cloud infrastructure. Indeed, they do not require much computing power and can significantly benefit from web-based visualization tools. Given that, we used StreamFlow to model the pipeline as a hybrid workflow, offloading the training portions to HPC nodes and collecting the resulting networks on the host execution flow for visualization purposes. In particular, different portions of the training spectrum have been offloaded to three different heterogeneous architectures:

- The ENEA CRESCO D.A.V.I.D.E. cluster, composed of 45 nodes with 2 IBM POWER8 sockets, 256 GB of RAM and 4 NVidia P100-SMX2 GPUs each, that can be used on-demand through bare SSH connections;

- The CINECA MARCONI100 cluster, a SLURM-managed HPC facility with 32 IBM POWER9 cores, 256 GB of RAM and 4 NVidia V100 GPUs per node;
- The High-Performance Computing for Artificial Intelligence (HPC4AI) infrastructure at the University of Torino, a multi-tenant hybrid Cloud-HPC system with 80 cores and 4 GPUs per node (T4 or V100-SMX2) managed by OpenStack [7].

As an interface towards Cloud-HPC infrastructures, StreamFlow seamlessly manages data movements and remote step execution with each of these infrastructures, automatically transferring back the training results to the Cloud-based host node to perform post-training steps.

## 5 Conclusion and future work

Presenting the work on the universal COVID-19 pipeline, we demonstrated that AI can be an effective support for human activity, that HPC is crucial to perform complex tasks in a useful time, and last but not least, that the convergence of different platforms is the next big thing. In this scenario, the general adoption of hybrid infrastructures from the scientific communities can only be obtained by leveraging advanced software tools like StreamFlow to enable portability and reproducibility.

The computing power required by the largest AI training runs has been increasing exponentially with a 3.4-month doubling time in the last 10 years [8]. A need that today can be only matched by way of accelerated computing provided by specialized processors such as GPUs and TPUs. With this background, the next era of HPC will inevitably see a further increase of heterogeneous hardware, with general-purpose CPUs flanked by highly parallel co-processors as GPUs and special-purpose hardware as neuromorphic chips and quantum annealing. Even if each architecture comes with its peculiar programming paradigm for local computations, the accelerator pattern is becoming a de-facto standard for moving computation away from CPUs.

We believe that a sound and stable system software part is crucial for the mainstream industrial adoption of HPC, enabling technology to transform applications into easily usable services hence into innovation. While in scientific computing the modernization of HPC applications is a scientific desideratum required to boost industrial adoption, the shift toward the Cloud model of services is a must in AI. AI applications are already modern, and they will not step back.

We advocate StreamFlow as an intuitive programming paradigm to foster the design of portable and scalable AI pipelines and reduce technical barriers to HPC facilities for domain experts without a strong computer science background.

Such paradigms can be further extended and improved. From one side, support for specific hardware (e.g., quantum processors) can be added to the list of connectors offered by hybrid WMSs. Moreover, more intuitive and user-friendly technologies (e.g., Jupyter Notebook) can be augmented with hybrid workflow semantics to evolve them from prototyping technologies to production-ready toolchains. Both these challenges are essential parts of the StreamFlow roadmap, together with further applications in the domains of deep learning, bioinformatics and molecular dynamics simulations.

## Funding & Acknowledgement

We gratefully acknowledge the support of Francesco Iannone from ENEA and the CRESCO/ENEAGRID High Performance Computing infrastructure and its staff. This work has been partially supported by the DeepHealth project, which has received funding from the European Union's Horizon 2020 research and innovation programme under grant agreement No. 825111, by the HPC4AI project funded by Regione Piemonte (POR FESR 2014-20 - INFRA-P). Access to CINECA resources has been possible thanks to the CINI-CLAIRE-ABD MoU 2020 supporting research on COVID-19.

We want to thank Emanuela Girardi and Gianluca Bontempi, who are coordinating the CLAIRE task force on COVID-19, for their support, and the group of volunteer researchers who contributed to the development of the CLAIRE COVID-19 universal pipeline, they are: Marco Calandri and Piero Fariselli (Radiomics & medical science, University of Torino, Italy); Marco Grangetto, Enzo Tartaglione (Digital image processing Lab, University of Torino, Italy); Simone Palazzo, Isaak Kavasidis (PeRCeiVe Lab, University of Catania, Italy); Bogdan Ionescu, Gabriel Constantin (Multimedia Lab @ CAMPUS Research Institute, University Politechnica of Bucharest, Romania); Miquel Perello Nieto (Computer Science, University of Bristol, UK); Inês Domingues (School of Sciences University of Porto, Portugal).